\documentclass[aps,prl,twocolumn,showpacs,showkeys,footinbib,superscriptaddress]{revtex4-1}

\usepackage{amsmath}
\usepackage{amssymb}
\usepackage{graphicx}
\usepackage{epstopdf}
\usepackage{bm}
\usepackage{color}
\usepackage{relsize}
\usepackage{braket}
\usepackage[caption=false]{subfig}

\usepackage{hyperref}
\usepackage[all]{hypcap}

\renewcommand{\i}{\ensuremath{\mathrm{i}}}
\newcommand{\e}{\ensuremath{\mathrm{e}}}
\renewcommand{\d}{\ensuremath{\mathrm{d}}}

\newcommand{\sgn}{\operatorname{{\mathrm sgn}}}

\begin{document}
\title{Tunneling Planar Hall Effect in Topological Insulators: Spin-Valves and Amplifiers}

\author{Benedikt Scharf}
\affiliation{Department of Physics, University at Buffalo, State University of New York, Buffalo, NY 14260, USA}
\author{Alex Matos-Abiague}
\affiliation{Department of Physics, University at Buffalo, State University of New York, Buffalo, NY 14260, USA}
\author{Jong E. Han}
\affiliation{Department of Physics, University at Buffalo, State University of New York, Buffalo, NY 14260, USA}
\author{Ewelina M. Hankiewicz}
\affiliation{Institute for Theoretical Physics and Astrophysics, University of W\"{u}rzburg, Am Hubland, 97074 W\"{u}rzburg, Germany}
\author{Igor \v{Z}uti\'c}
\affiliation{Department of Physics, University at Buffalo, State University of New York, Buffalo, NY 14260, USA}

\date{\today}

\begin{abstract}
We investigate tunneling across a single ferromagnetic barrier on the surface of a three-dimensional topological insulator. In the presence of a magnetization component along the bias direction, a tunneling planar Hall conductance (TPHC), transverse to the applied bias, develops. Electrostatic control of the barrier enables a giant Hall angle, with the TPHC exceeding the longitudinal tunneling conductance. By changing the in-plane magnetization direction it is possible to change the sign of both the longitudinal and transverse differential conductance without opening a gap in the topological surface state. The transport in a topological insulator/ferromagnet junction can thus be drastically altered from a simple spin-valve to an amplifier.
\end{abstract}

\keywords{}

\maketitle

Exotic properties of three-dimensional topological insulators (3D TIs) arise from their helical surface states, described as 2D Dirac fermions with spin-momentum locking~\cite{Hasan2010:RMP,*Qi2011:RMP,*Shen2012}. Topological insulators have large spin-orbit coupling (SOC) leading to striking manifestations of the conservation of angular momentum from a colossal Kerr rotation~\cite{Aguilar2012:PRL} and photocurrent control~\cite{McIver2015:NN} to magnetization switching~\cite{Mellnik2014:N,*Fan2014:NM}. The interplay between magnetism and SOC in ferromagnet(F)/TI junctions provides a versatile platform to study fundamental effects and spintronic applications~\cite{Hasan2010:RMP,*Qi2011:RMP,*Shen2012,Mellnik2014:N,*Fan2014:NM}. Previous tunneling studies have largely focused on the longitudinal response~\cite{Mondal2010:PRL,Wu2010:PRB3,*Wu2012:NRL,Li2014:PRB,*Li2014:NN,Tian2014:SSC,*Tian2015:NSR} since a common expectation in tunnel junctions is that the transverse (Hall) response is negligible.

\begin{figure}[t]
\centering
\includegraphics*[trim=0cm 9.0cm 3.1cm 8.0cm,clip,width=8cm]{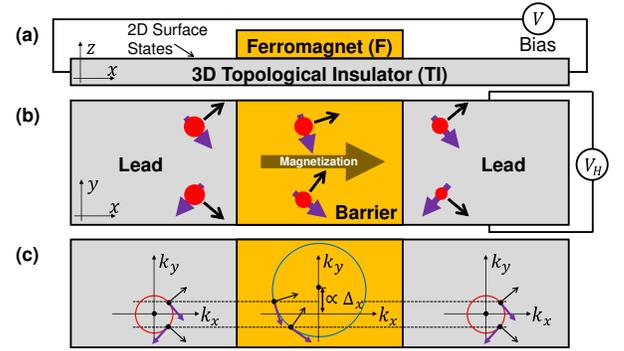}
\caption{(a) Schematic setup. (b) Origin of the planar Hall conductance and net Hall voltage, $V_H$, due to asymmetric tunneling. The circle sizes represent the asymmetry in transmission probabilities arising from the interfacial mismatch of spin directions (locked to the velocity). (c) Spin mismatch: Fermi circles in the TI (upper Dirac cone) and the barrier (lower Dirac cone, shifted by a proximity-induced exchange splitting $\Delta_x$). In~(b) and~(c) violet (black) arrows denote the electron spin orientation (direction of motion).}
\label{fig:Scheme}
\end{figure}

In contrast to previous manifestations of the Hall effect, such as the anomalous~\cite{Sinitsyn2008:JPCM,*Nagaosa2010:RMP,Culcer2011:PRB}, tunneling anomalous~\cite{Vedyayev2013:PRL,*Vedyayev2013:APL,Tarasenko2004:PRL,MatosAbiague2015:PRL,*Dang2015:PRB}, and planar Hall effects~\cite{Tang2003:PRL,*Seemann2011:PRL}, we propose an unexplored tunneling planar Hall effect (TPHE) emerging in F/TI junctions (Fig.~\ref{fig:Scheme}), qualitatively different from these manifestations in terms of the relevant geometry and the magnetization configuration. In particular, the proposed effect is maximized for a planar magnetization parallel to the applied bias, where other Hall effects vanish~\cite{SM}.

Unlike in conventional tunneling, a thick barrier with TIs can still lead to a large conductance due to Klein tunneling~\cite{SM}. We show that an asymmetry in the tunneling conductance due to the in-plane barrier magnetization enables efficient transverse (Hall) spin-valves. With spin-momentum locking and a tunable resonant transmission, these spin-valves can display a {\em transverse} negative differential (ND) conductance even in the limit of vanishing applied bias, suggesting a path to amplifiers and other active spintronic devices~\cite{Zutic2004:RMP,*Fabian2007:APS}.

This peculiar behavior arises from asymmetric tunneling of electrons with opposite incident angles through the barrier [Fig.~\ref{fig:Scheme}(b)]. The finite tunneling planar Hall conductance (TPHC) can be understood as the spin mismatch between TI and F selecting electrons with positive transverse velocity~\footnote{In Fig.~\ref{fig:Scheme}, we discuss the regime of $V_0>\varepsilon_\mathrm{F}>0$. Our arguments, however, are also valid for other regimes.} to be transmitted more effectively [Fig.~\ref{fig:Scheme}(b)]. The interfacial spin mismatch results from spin-momentum locking and a shift of the Dirac cone due to the exchange splitting [Fig.~\ref{fig:Scheme}(c)]. Translational symmetry along the $y$-axis yields an effective Snell's law~\cite{Zutic2000:PRB} preserving the transverse momentum, while the longitudinal momentum changes sign on the lower Dirac cone (the group velocity points to its apex, see Ref.~\onlinecite{SM}).

Our system is described by the effective Hamiltonian
\begin{equation}\label{Complete_Hamiltonian}
\hat{H}_0=v_\mathsmaller{\mathrm{F}}\left(\bm{\sigma}\times\hat{\bm{p}}\right)\cdot\bm{e}_z+\left(V_0-\bm{\Delta}\cdot\bm{\sigma}\right)h(x)
\end{equation}
with the barrier function $h(x)=\Theta(-x)\Theta(x+d)$ for a square (finite) barrier of width $d$ and $h(x)=d\delta(x)$ for the respective $\delta$-barrier. Here, $v_\mathsmaller{\mathrm{F}}$ is the Fermi velocity of the surface states ($v_\mathsmaller{\mathrm{F}}\approx6\times10^5$ m/s in Bi$_2$Se$_3$~\cite{Zhang2009:NP,*Kim2012:NP}), $\hat{\bm{p}}$ and $\bm{\sigma}$ denote vectors containing the momentum operators and Pauli spin matrices~\cite{Hasan2010:RMP,*Qi2011:RMP,*Shen2012}, while $\bm{\Delta}$ and $V_0$ describe the proximity-induced ferromagnetic exchange splitting and an electrostatic potential barrier, respectively. A planar exchange field $\bm{\Delta}$ shifts the apex of the Dirac cones from the origin to $(-\Delta_y/\hbar v_\mathsmaller{\mathrm{F}},\Delta_x/\hbar v_\mathsmaller{\mathrm{F}})^T$ in the $k_xk_y$-plane. Therefore, for $\Delta_y =0$ the longitudinal (transverse) transport is even (odd) in $\Delta_x$. In Eq.~(\ref{Complete_Hamiltonian}), we focus on F/TI junctions where the topological surface states (TSSs) are decoupled from bulk states~\cite{SM}.

The conductance for a bias along the $x$-direction is obtained from the eigenstates of Eq.~(\ref{Complete_Hamiltonian}) with energy $E$ and conserved momentum $\hbar k_y$ [Fig.~\ref{fig:Scheme}(c)], $\Psi_{k_y}(x,y)=\exp(\i k_yy)\Phi(x)/\sqrt{2S}$ with the surface area $S$ and
\begin{equation}\label{FiniteBarrier_States}
\Phi(x)=\left\{\begin{array}{ll}
 \chi_\mathsmaller{+}\e^{\i k_xx}+r_\mathrm{e}\chi_\mathsmaller{-}\e^{-\i k_xx}, & x<-d,\\
 l\tilde{\chi}_\mathsmaller{+}\e^{\i\tilde{k}_\mathsmaller{+}x}+m\tilde{\chi}_\mathsmaller{-}\e^{\i\tilde{k}_\mathsmaller{-}x}, & -d<x<0,\\
 t_\mathrm{e}\chi_\mathsmaller{+}\e^{\i k_xx}, & x>0\\
 \end{array}\right.
\end{equation}
for the finite barrier. For the $\delta$-barrier, the states $\Phi(x<0)$ and $\Phi(x>0)$ are given by the first and third lines of Eq.~(\ref{FiniteBarrier_States}), respectively. Defining the angle $-\pi/2\leq\theta\leq\pi/2$ as $\hbar v_\mathsmaller{\mathrm{F}}k_x=|E|\cos\theta$ and $\hbar v_\mathsmaller{\mathrm{F}}k_y=|E|\sin\theta$, the momenta are given by $\hbar v_\mathsmaller{\mathrm{F}}\tilde{k}_\pm=-\Delta_y\pm\hbar v_\mathsmaller{\mathrm{F}}\tilde{k}_x$ and the spinors by $\chi_\mathsmaller{\pm}=(1,b_\mathsmaller{\pm})^T$ and $\tilde{\chi}_\mathsmaller{\pm}=(1,\tilde{b}_\mathsmaller{\pm})^T$ with $b_\mathsmaller{\pm}=\mp\i\sgn(E)\e^{\pm\i\theta}$, $\tilde{b}_\mathsmaller{\pm}=\left[\left(|E|\sin\theta-\Delta_x\right)\mp\i\hbar v_\mathsmaller{\mathrm{F}}\tilde{k}_x\right]/\left(E-V_0-\Delta_z\right)$, and
\begin{equation}\label{FiniteBarrier_Momentum}
\hbar v_\mathsmaller{\mathrm{F}}\tilde{k}_x(E,\theta)=\sqrt{\left(E-V_0\right)^2-(\Delta_x-|E|\sin\theta)^2-\Delta^2_z}.
\end{equation}

Carefully invoking the boundary conditions~\cite{MatosAbiague2003:PRA,SM} to determine $r_\mathrm{e}$, $t_\mathrm{e}$, $l$, $m$ in Eq.~(\ref{FiniteBarrier_States}) yields the transmission
\begin{equation}\label{Transmission}
\begin{aligned}[l]
T(E,\theta)=\frac{1}{1+\frac{\left(V_0\sgn(E)\sin\theta-\Delta_x\right)^2+\Delta^2_z\cos^2\theta}{\left(\hbar v_\mathsmaller{\mathrm{F}}/d\right)^2\cos^2\theta}\frac{\sin^2Z_\mathrm{eff}}{Z^2_\mathrm{eff}}},
\end{aligned}
\end{equation}
where $Z_\mathrm{eff}=\tilde{k}_x(E,\theta)d$ for a finite barrier and $Z_\mathrm{eff}=\sqrt{V_0^2-\Delta^2}d/(\hbar v_\mathsmaller{\mathrm{F}})$ for a $\delta$-barrier with $\Delta=\sqrt{\Delta^2_x+\Delta^2_z}$. Here, $T(E,\theta)$ is independent of $\Delta_y$ and asymmetric with respect to $\theta$ for finite $\Delta_x$.

We focus on the case $\Delta=|\Delta_x|$, $\Delta_z=0$, while the effects of finite $\Delta_z$ are discussed in Ref.~\onlinecite{SM}. The transmission from Eq.~(\ref{Transmission}) displays two qualitatively different regimes: (i) oscillatory, with real $Z_\mathrm{eff}$ as a consequence of Klein tunneling in Dirac systems like graphene~\cite{Katsnelson2006:NP,*Bhattacharjee2006:PRL,*Linder2007:PRL}, and (ii) decaying, with complex $Z_\mathrm{eff}$ and typical for massive low-energy systems described by Schr\"{o}dinger's equation. A remarkable property of our system is that by controlling the magnetization and/or the top gate potential (recall $Z_\mathrm{eff}$ depends on $V_0$ and $\Delta$) it is possible to switch between the two regimes and produce very large differences in $T(E,\theta)$.

Such a tunable transmission can lead to a large anisotropy for some incident angles. In the oscillatory regime, in particular, we find from Eq.~(\ref{Transmission}) that perfect transmission is realized for
\begin{equation}\label{Resonances}
V_0\sgn(E)\sin\theta=\Delta\;\;\textrm{or}\;\;Z_\mathrm{eff}(E,\theta)=n\pi,\;\; n=1,2,...\quad.
\end{equation} 
Here, the first equality describes perfect transmission at each interface due to the absence of any spin mismatch between TIs and F. The second equality is a resonance condition for constructive interference when a multiple of the longitudinal wavelength $2\pi/\tilde{k}_x$ matches $d$~\footnote{While the first condition is only valid for $\Delta_z=0$ and $|V_0|>|\Delta_x|$, the second condition given by Eq.~\ref{Resonances} is also valid for $\Delta_z\neq0$.}.

Using Eq.~(\ref{Transmission}), the conductance at zero temperature, for a bias applied in the $x$-direction, reads as~\cite{SM}
\begin{equation}\label{Conductance_Gxxyx}
G_{xx/yx}=\frac{e^2}{h}\frac{|\varepsilon_\mathrm{F}|D_{x/y}}{2\pi\hbar v_\mathsmaller{\mathrm{F}}}\int\limits_{-\pi/2}^{\pi/2}\d\theta\;T(\varepsilon_\mathrm{F},\theta)
\left\{\begin{array}{l}\cos\theta\\ \sgn(\varepsilon_\mathrm{F})\sin\theta\\ \end{array}\right.,
\end{equation}
where $D_{x/y}$ is the width perpendicular to the current flow in the $x/y$-direction and -$e$ is the electron charge. We normalize $G_{xx/yx}$ to the Sharvin conductance (transparent barrier), $G_{0x/y}=\left(e^2/h\right)\left|\varepsilon_\mathrm{F}\right|D_{x/y}/\left(\pi\hbar v_\mathsmaller{\mathrm{F}}\right)$~\cite{Zutic1999:PRB}.

\begin{figure}[t]
\centering
\includegraphics*[trim=0cm 6.5cm 0cm 6.25cm,clip,width=8.6cm]{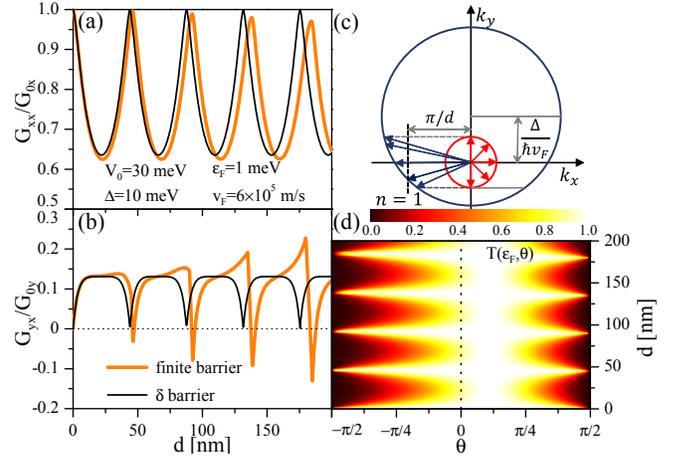}
\caption{Dependence of the (a) longitudinal and (b) transverse conductances on $d$ for finite and $\delta$-barriers. (c) Fermi circles in the leads (inner circle) and barrier (outer circle). The arrows denote the wave vectors of states with positive $x$-component of the velocity and the vertical dashed line indicates the first-order resonance condition. (d) Transmission $T(\varepsilon_\mathrm{F},\theta)$ of a finite barrier as a function of $d$ and $\theta$.}\label{fig:Width}
\end{figure}

For a $\delta$-barrier and $|V_0|\gg\Delta$, Eq.~(\ref{Transmission}) can be expanded up to the lowest order in $\Delta/V_0$,
\begin{eqnarray}\label{Conductance_GxxLimit}
G_{xx}/G_{0x}&\approx&\sec^2Z_0-\mathrm{tanh}^{-1}\left|\cos Z_0\right|\tan^2 Z_0/\left|\cos Z_0\right|,\quad \\
G_{yx}/G_{0y}&\approx&(\pi\Delta/2V_0)\left|\sin Z_0\right|\left(1-\left|\sin Z_0\right|\right)^2/\cos^4Z_0,\quad \label{Conductance_GyxLimit}
\end{eqnarray}
where $Z_0=V_0d/(\hbar v_\mathsmaller{\mathrm{F}})$ \footnote{For $\Delta_z\neq0$, $\Delta$ is replaced by $\Delta_x$ in Eq.~(\ref{Conductance_GyxLimit}).}. These expressions capture the oscillatory behavior of $G_{xx/yx}$ and reveal that at the resonance condition, $Z_\mathrm{eff}\approx Z_0=n\pi$, $G_{xx}=G_{0x}$ reaches perfect transmission, whereas $G_{yx}$ vanishes. Such a qualitative behavior is corroborated by the full $\delta$-barrier dependence of $G_{xx/yx}$ on $d$, shown in Figs.~\ref{fig:Width}(a) and~(b). Even though the $\delta$-barrier provides a good approximation for small $d$, it fails to describe the appearance of $G_{yx}<0$ and the increase of its amplitude with $d$. Hence, we will focus on the finite barrier and employ the $\delta$-model only to obtain analytical approximations.

The main features observed in Figs.~\ref{fig:Width}(a) and~(b) can be understood by analyzing the phase space available for tunneling shown in Fig.~\ref{fig:Width}(c) for $V_0>\varepsilon_\mathrm{F}>0$. Here, the inner (outer) circle with radius $|\varepsilon_\mathrm{F}|/(\hbar v_\mathsmaller{\mathrm{F}})$ [$|V_0-\varepsilon_\mathrm{F}|/(\hbar v_\mathsmaller{\mathrm{F}})$] represents the $k$-space Fermi circle in the leads (barrier) and the arrows indicate the Fermi wave vectors of the scattering states available for transport. As discussed in Fig.~\ref{fig:Scheme}, the asymmetry between the scattering states with $k_y>0$ ($0<\theta<\pi/2$) and $k_y<0$ ($-\pi/2<\theta<0$) due to $\Delta$ causes a finite TPHC. For illustration, we show in Fig.~\ref{fig:Width}(d) the transmission, $T(\varepsilon_\mathrm{F},\theta)$, of a finite barrier as a function of $d$ and $\theta$. The asymmetry of $T(\varepsilon_\mathrm{F},\theta)$ with respect to $\theta=0$ due to the first equality in Eq.~(\ref{Resonances}) can clearly be seen, which results in the appearance of a nonzero $G_{yx}$ after the integration in Eq.~(\ref{Conductance_Gxxyx}). On the other hand, the oscillatory behavior with $d$ in Fig.~\ref{fig:Width}(d) is governed by $\sin^2Z_\mathrm{eff}$ in Eq.~(\ref{Transmission}).

When $|V_0-\varepsilon_\mathrm{F}|>\Delta+|\varepsilon_\mathrm{F}|$, the Fermi circle of the leads is inside that of the barrier as shown in Fig.~\ref{fig:Width}(c). Then, for each Fermi vector in the leads, there is one available in the barrier and the system is purely in the Klein tunneling regime. The deviations between the finite and $\delta$-barrier models with increasing $d$ originate from the angular dependence of $Z_\mathrm{eff}$ and the ensuing asymmetric resonances in the case of a finite barrier, explained by Fig.~\ref{fig:Width}(c): With increasing $d$, the first-order resonance [$n=1$ in Eq.~(\ref{Resonances})] moves towards smaller $k_x$-values and, at $d\approx46$ nm, it crosses the Fermi circle of the barrier. The first states reaching the resonance are those with $k_y>0$, causing an increase in $G_{yx}$ compared to the $\delta$-barrier model. As $d$ is further increased, the resonance moves to states with $k_y<0$ producing a fast decrease in $G_{yx}$, which, eventually, becomes negative. In thicker barriers, the trend repeats periodically with $d$ each time a new resonance becomes relevant. This occurrence of multiple resonances ($n=1,2$, etc) results in the increase of the amplitude of the TPHC for even larger values of $d$ (if $|\varepsilon_\mathsmaller{\mathrm{F}}|\ll |V_0|$) as shown in Fig.~\ref{fig:Width}(b).

\begin{figure}[t]
\includegraphics*[trim=0cm 6.5cm 0cm 6.25cm,clip,width=8.6cm]{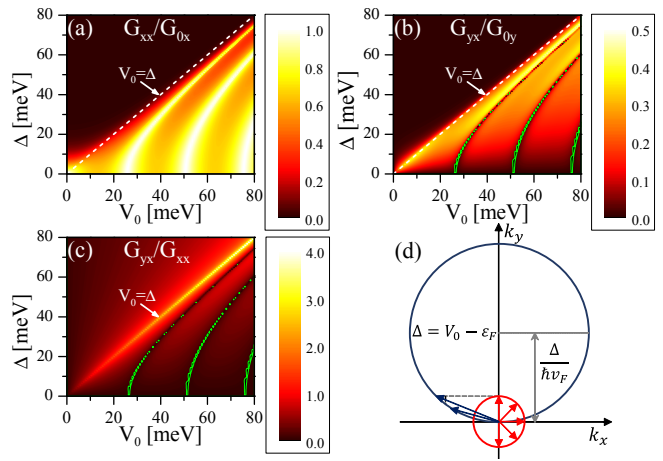}
\caption{Dependence of the (a) longitudinal and (b) transverse conductances as well as (c) of their ratio on $V_0$ and $\Delta$ for a finite barrier with $d=50$ nm, $\varepsilon_\mathrm{F}=1$ meV, and $v_\mathrm{F}=6.0\times10^5$ m/s. Green lines: boundaries of regions with negative conductance. (d) Same as in Fig.~\ref{fig:Width}(c), but for larger $\Delta$.}\label{fig:DensityV0xc50nm}
\end{figure}

The interplay between $V_0$ and $\Delta$ and the appearance of a TPHC are illustrated by Fig.~\ref{fig:DensityV0xc50nm} for (a) $G_{xx}$, (b) $G_{yx}$, and (c) their ratio for a finite barrier with $d=50$ nm and a fixed $\varepsilon_\mathsmaller{\mathrm{F}}$. Figures~\ref{fig:DensityV0xc50nm}(a) and~(b) clearly show the transition from a region of oscillatory Klein tunneling ($|V_0|>\Delta+2\varepsilon_\mathrm{F}\approx\Delta$) to a region of decaying tunneling ($|V_0|<\Delta$). Such a transition can be understood by resorting to the analysis of the Fermi circles. As discussed above, the scheme in Fig.~\ref{fig:Width}(b) corresponds to the Klein tunneling regime, but increasing $\Delta$ will shift up the Fermi circle of the barrier, which at $\Delta=V_0-2\varepsilon_\mathrm{F}$ starts to cross the Fermi circle of the leads. Therefore, increasing $\Delta$ above that value results in the formation of an intermediate regime in which only a part of the available states can undergo Klein tunneling, while the other experiences decaying tunneling. The contrast between the two tunneling mechanisms becomes extreme when $\Delta=V_0-\varepsilon_\mathrm{F}$. In such a situation, as shown in Fig.~\ref{fig:DensityV0xc50nm}(d), Klein tunneling occurs only for states with $k_y>0$, while those with $k_y<0$ undergo decaying tunneling. This strong asymmetry in the tunneling favors the transmission of states with larger $k_y$ values and results in a remarkably large ratio between the TPHC and the longitudinal conductance. As shown in Fig.~\ref{fig:DensityV0xc50nm}(c), such a ratio can even exceed 1, implying large Hall angles, $\theta_{\rm H}=\arctan(G_{yx}/G_{xx})\approx 75^\circ$ for the parameters chosen here. Such giant values of the Hall angle are comparable to those recently detected in a 3D magnetic TI~\cite{Kandala2015:NC}. Green lines in Figs.~\ref{fig:DensityV0xc50nm}(b) and~(c) indicate negative values of the TPHC, whose origin is the same as in Fig.~\ref{fig:Width}(c).

The $\delta$-barrier model enables us to obtain an analytical expression for the giant Hall angle. Indeed, for $|V_0|\approx\Delta$,
\begin{equation}\label{Conductance_Ratio1}
\tan\theta_{\rm H}=\frac{G_{yx}}{G_{xx}}=\frac{\pi|Z_0|\left(|Z_0|-1\right)^2}{2\left[\left(2\mathrm{ln}|Z_0|-1\right)Z_0^2+1\right]},
\end{equation}
which increases with $|Z_0|$, even though $G_{xx}$ and $G_{yx}$ individually decrease (we assume $D_x=D_y$).

\begin{figure}[t]
\centering
\includegraphics*[trim=0cm 6.5cm 0cm 6.25cm,clip,width=8.6cm]{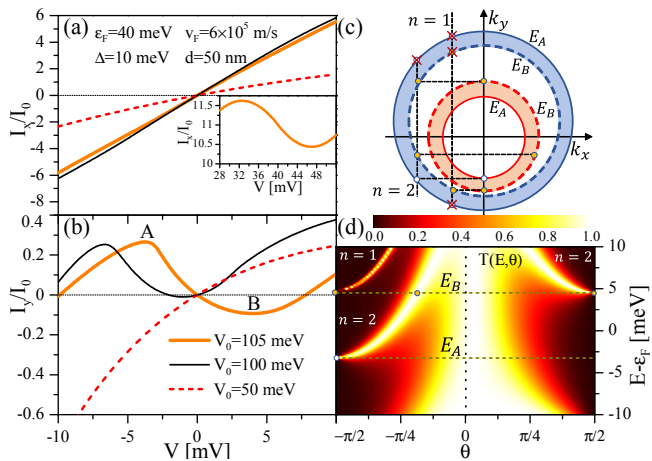}
\caption{Bias dependence of the (a) longitudinal and (b) transverse currents for a finite barrier and different $V_0$. Assuming $D_x=D_y=10$ $\mu$m, both currents are given in units of $I_0=12$ $\mu$A. (c) Same as in Fig.~\ref{fig:Width}(c), but with thickened Fermi circles accounting for a finite energy window around $\varepsilon_\mathrm{F}$. (d) $T(E,\theta)$ for $V_0=105$ meV. The inset in~(a) shows the appearance of a ND $G_{xx}$ for $V_0=105$ meV at high bias.}
\label{fig:Current}
\end{figure}

We next examine the current-voltage ($I$-$V$) characteristics and reveal the appearance of a negative differential (ND) conductance. While for $|V|\ll|\varepsilon_\mathrm{F}|$, Eq.~(\ref{Complete_Hamiltonian}) can be used to calculate the current, bias-induced changes of the electrostatic potential have to also be taken into account in general. For all $I$-$V$ calculations~\cite{SM}, we model this effect by adding the step-like~\cite{Tsu1973:APL} potential profile $V\left[\Theta(-x-d)+\Theta(-x)\right]/2$ to Eq.~(\ref{Complete_Hamiltonian}) and computing the transmitted currents for this system numerically. As a consequence, a ND longitudinal conductance observed in single barrier graphene transistors~\cite{Britnell2013:NC} appears also in our system [inset of Fig.~\ref{fig:Current}(a)]. Surprisingly, the transverse current, $I_y$, also shows a change of sign in its slope [segment from A to B in Fig.~\ref{fig:Current}(b)], the signature of a ND Hall conductance (NDHC),
even at low $V$ and within a range in which the differential longitudinal conductance remains positive [Fig.~\ref{fig:Current}(a)].

The appearance of a NDHC is exemplified for $V_0=105$ meV in Fig.~\ref{fig:Current}(b) with the corresponding transmission $T(E,\theta)$ displayed in Fig.~\ref{fig:Current}(d). Here, the key observation is that in the Klein tunneling regime, the asymmetry of the resonances with respect to $\theta=0$ depends on the energy. Indeed, as depicted in Fig.~\ref{fig:Current}(d), for different energies the resonances appear in the region $k_y<0$, or $k_y>0$, or in both. This behavior is explained in Fig.~\ref{fig:Current}(c), where the Fermi circles of the leads and barrier have been thickened to account for the energy window from $E_A$ (solid circles) to $E_B$ (dashed circles) around the Fermi energy, $\varepsilon_\mathrm{F}=40$ meV. The vertical lines marked by $n=1$ and $n=2$ indicate the resonance condition $Z_\mathrm{eff}(k_x,k_y)=n\pi$ as in Eq.~(\ref{Resonances}). Open and full (yellow) dots represent the resonances in Fig.~\ref{fig:Current}(d) at $E_A$ and $E_B$, while crossed dots represent resonances forbidden by the conservation of $k_y$. The nonmonotonic $I_y$-$V$ characteristic in Fig.~\ref{fig:Current}(b) follows from the positions of the resonances: The local maximum A emerges as the relevant energy window between $\varepsilon_\mathrm{F}$ and $\varepsilon_\mathrm{F}+eV$ starts to cross the resonance at $E_A$ for a $k_y<0$ [Figs.~\ref{fig:Current}(c) and~(d)] resulting in a reduced $I_y$ with $V$. This resonance is compensated for as another resonance favoring $k_y>0$ is reached at $E_B$ [Figs.~\ref{fig:Current}(c) and~(d)], giving rise to the local minimum B and subsequent increase of $I_y$ in Fig.~\ref{fig:Current}(b).

As shown in Fig.~\ref{fig:Current}(b), the NDHC present for $V_0=100$ meV and $V_0=105$ meV is suppressed at $V_0=50$ meV, suggesting the possibility of controlling the NDHC by gate-tuning the barrier. Moreover, the $I_y$-$V$ characteristic for $V_0=105$ meV resembles that of a typical active ND resistor, which is unusual for tunneling systems \footnote{Typically, resonant tunneling devices exhibit $I$-$V$ characteristics of passive ND resistors, where the ND conductance region does not cross the origin.}.

Despite the simplicity of a single ferromagnetic region, our system exhibits a variety of functionalities expected to require more complex spintronic devices~\cite{Maekawa2012,Wunderlich2010:S}. In addition to a spin-valve operation for magnetic sensing and storing information, shown in Figs.~\ref{fig:Current}(a) and~(b), positive, negative, and ND conductances can be tuned by properly adjusting the barrier potential, suitable for processing information. Such different behaviors in the same system are attractive for potential applications in reconfigurable devices operating as feedback oscillators, active filters, modulators, and amplifiers~\cite{Boylestad2012}. These functionalities can be alternated both by the barrier potential and in a nonvolatile way using the magnetization orientation.

Our findings, expressed using Bi$_2$Se$_3$ parameters, could also be detected in other, more suitable, TIs to avoid the coexistence of bulk and TSSs at the Fermi level, even after adding a magnetic region~\cite{Eremeev2013:PRB,Lee2014:PRB2,Chang2015:PRB}. Alloying can help to tune the Fermi level inside the bulk bandgap in (Bi,Sb)$_2$Te$_3$, (Bi$_2$,Sb)(Te$_3$,Se), or Tl(Bi,Sb)Te$_2$~\cite{Chen2010:S,Arakane2012:NC,Kushwaha2016:NC,Trang2016:PRB,Weyrich2015:arxiv,*Ando2014:NL}, while gating strained HgTe or capped Bi$_2$Te$_3$ can isolate TSSs~\cite{Bruene2014:PRX,Ngabonziza2016:AEM}. Recent experiments imply a dominant role of TSSs in junctions with magnetic regions, such as YIG/(Bi,Sb)$_2$Te$_3$ with an independent tuning of electronic properties and proximity-induced magnetism in TIs~\cite{Jiang2015:NL,*Jiang2016:NC}. Magnetic proximity effects have been observed even at 300 K in EuS/Bi$_2$Se$_3$ or (Bi,Mn)Te~\cite{Katmis2016:N,Vobornik2011:NL}.

To realize magnetic proximity effects for the in-plane transport, magnetic insulators are desirable. This precludes current flow in the more resistive F region [Fig.~\ref{fig:Scheme}(a)] and minimizes hybridization with the TI to enable a gate-tunable proximity-induced exchange splitting in the surface states. However, as shown by the example of tunable magnetic proximity effects in graphene~\cite{Lazic2016:PRB}, one could instead employ ferromagnetic metals, separated by an insulating region from the TI. In graphene, the interplay of proximity induced SOC and magnetism can also yield interesting effects~\cite{Lee2016:arxiv}.

Even in the presence of additional states, such as Rashba 2D states, a finite TPHE can still be expected. Those states will, in general, also exhibit a spin mismatch and thus contribute to the transverse Hall voltage, potentially competing with the TSSs~\cite{SM}. Nevertheless, experiments on current-induced spin polarization, suggest that these two contributions are inequivalent and their relative significance can be tuned by changing the position of the Fermi level~\cite{Liu2015:PRB,Yang2016:PRB,Li2016:arxiv}. Future work could involve complementary first-principles transport studies to quantify the influence of additional topologically trivial states and studying the role of phonons, shown to profoundly affect transport in TIs~\cite{Costache2014:PRL}.

\noindent\emph{Acknowledgments.} We thank Y. Ando, K. Belashchenko, J. Friedman, A. Khitun, L. Molenkamp, K. Park, T. Valla, and J. Moodera for valuable discussions. This work was supported by U.S. DOE, Office of Science BES, under Award DESC0004890 (A.M.-A., I.\v{Z}.), by by U.S. ONR Grant No. N000141310754, NSF-ECCS1508873 (B.S.), the German Science Foundation (DFG) Grant No. SCHA 1899/1-1 (B.S.), and DFG Grant No. HA 5893/4-1 within SPP 1666 (E.M.H.).

\bibliography{BibTopInsAndTopSup}

\section{Comparison between different Hall effects}\label{Sec:HallEffects}

A central objective for spintronic device applications is to control electron transport by utilizing its spin~\cite{Zutic2004:RMP,*Fabian2007:APS}. In this context, the anomalous~\cite{Sinitsyn2008:JPCM,*Nagaosa2010:RMP} and planar~\cite{Tang2003:PRL,*Seemann2011:PRL} Hall effects (AHE and PHE), both arising from the interplay between magnetism and spin-orbit coupling (SOC), offer intriguing possibilities for devices such as Hall sensors. Recent studies also indicate the existence of a tunneling AHE (TAHE), where the combination of tunneling, interfacial SOC, and magnetism just by themselves has been shown to lead to a finite TAHE~\cite{Vedyayev2013:PRL,*Vedyayev2013:APL,MatosAbiague2015:PRL,Dang2015:PRB}. Conversely, we predict a tunneling PHE (TPHE) in electrostatically gated ferromagnetic junctions based on three-dimensional topological insulators (3D TIs).

\begin{figure}[t]
\centering
\includegraphics*[trim=0cm 8.0cm 1.0cm 8.0cm,clip,width=8.6cm]{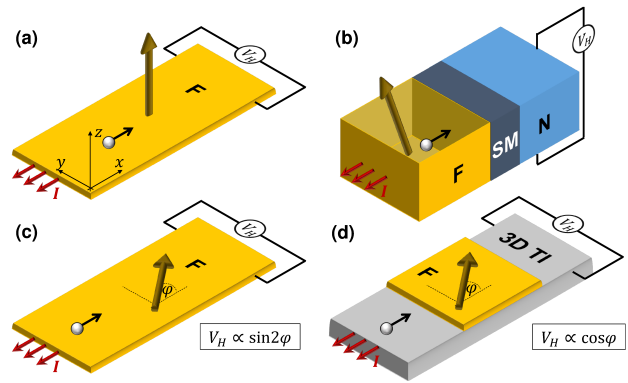}
\caption{(Color online) Schematic setups of the (a) anomalous, (b) tunneling anomalous, (c) planar and (d) tunneling planar Hall effects. Here, thick brown arrows denote the magnetization, black arrows the direction of electron motion, and red arrows the direction of current flow $\bm{I}$. In each case, a Hall voltage $V_H$ emerges perpendicular to the direction of current flow/bias direction. Ferromagnetic, normal, and semiconducting regions are labeled as F, N, and SM, respectively, and topological insulators are labeled as TI.}\label{fig:SuppHallEffects}
\end{figure}

Schematic setups of those four effects are shown in Fig.~\ref{fig:SuppHallEffects}. While all of them emerge due to SOC and magnetism, the underlying mechanisms are notably different as are their respective geometries and magnetization configurations. The AHE arises in ferromagnetic materials, either intrinsically due to a finite net anomalous velocity contribution from all occupied bands or extrinsically due to spin-dependent scattering at impurities, and is determined by the out-of-plane magnetization [see Fig.~\ref{fig:SuppHallEffects}~(a)]. Here, the bias direction and the magnetization component causing the AHE are perpendicular to each other. 
Similarly, the TAHE emerges in ferromagnet (F)/semiconductor (SM)/normal (N)~\cite{MatosAbiague2015:PRL} or other magnetic tunnel junctions~\cite{Dang2015:PRB} due to a magnetization (and interfacial spin-orbit fields) perpendicular to the bias direction [see Fig.~\ref{fig:SuppHallEffects}~(b)]. Figure~\ref{fig:SuppHallEffects}~(c), on the other hand, illustrates that the PHE arises in ferromagnetic materials due to an in-plane magnetization with a Hall signal $V_H\propto\sin2\varphi$, where $\varphi$ is the angle between the direction of current flow and the in-plane magnetization.

Like the PHE, the TPHE in TIs proposed in this work is caused by an in-plane magnetization. Equation~(8) in the main text, however, implies---up to the lowest order---a Hall signal $V_H\propto\cos\varphi$. Thus, whereas the PHE vanishes if the magnetization is oriented along the bias direction, $\varphi=0$, such a configuration is ideal for a giant TPHE. Moreover, the TPHE, like the TAHE, but unlike the AHE and the PHE, can emerge in a nonmagnetic region of a tunnel junction. For the TAHE, however, at least one of the leads is required to be magnetic and provide a source of spin polarization, whilst the TPHE appears even in the absence of any magnetic leads. Hence, despite similarities, the TPHE in TIs is a new effect due to an interfacial spin mismatch distinct from the PHE, AHE, and TAHE.

\section{Linear dispersion, spin-momentum/velocity locking, and Klein tunneling}\label{Sec:AppendixKlein}

\begin{figure}[t]
\centering
\includegraphics*[trim=0cm 8.0cm 1.0cm 8.0cm,clip,width=8.6cm]{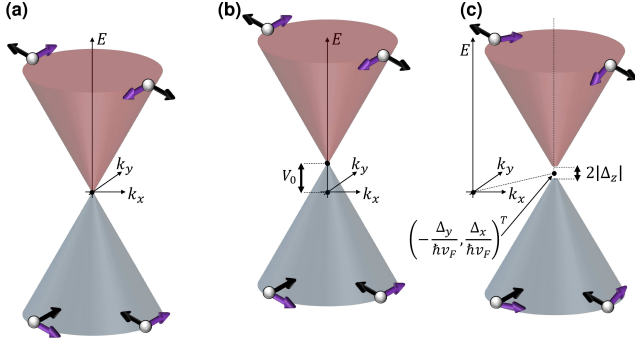}
\caption{(Color online) Low-energy excitation spectra given by Eq.~(\ref{Dispersion}) for (a) 3D TI surface states, (b) 3D TI surface states in the presence of an electrostatic potential $V_0$, and (c) 3D TI surface states in the presence of a ferromagnetic exchange splitting $\bm{\Delta}=(\Delta_x,\Delta_y,\Delta_z)^T$. Here, the spin and velocity expectation values of eigenstates of the system with momentum $\hbar\bm{k}=(\hbar k_x,\hbar k_y)^T$ are denoted by violet and black arrows, respectively.}\label{fig:SuppDiracCones}
\end{figure}

The surface states of 3D TIs are described by a 2D Dirac-like Hamiltonian, which in the presence of an electrostatic potential $V_0$ and an induced ferromagnetic exchange field $\bm{\Delta}=|\bm{\Delta}|\bm{n}\equiv\Delta(n_x,n_y,n_z)^T\equiv(\Delta_x,\Delta_y,\Delta_z)^T$ with the magnetization orientation of the ferromagnet on top the TI surface given by the unit vector $\bm{n}=(n_x,n_y,n_z)^T$ yields the dispersion
\begin{equation}\label{Dispersion}
E=V_0\pm\sqrt{\left(\hbar v_\mathsmaller{\mathrm{F}}k_x+\Delta_y\right)^2+\left(\hbar v_\mathsmaller{\mathrm{F}}k_y-\Delta_x\right)^2+\Delta_z^2}.
\end{equation}
Here, hybridization effects between opposite surfaces have been neglected, which would give an additional contribution to the gap between conduction and valence states. 

In the absence of any external field, $V_0=0$ and $\bm{\Delta}=\bm{0}$, Eq.~(\ref{Dispersion}) exhibits a linear dispersion as illustrated by the Dirac cones in Fig.~\ref{fig:SuppDiracCones}~(a). Similarly, a finite electrostatic potential $V_0$, as shown in Fig.~\ref{fig:SuppDiracCones}~(b), preserves the Dirac cones, but shifts their energies. A finite exchange splitting, on the other hand, can have a more profound effect on the shape of the dispersion [see Fig.~\ref{fig:SuppDiracCones}~(c)]: (i) Components of the magnetization direction parallel to the TI surface, $\Delta_x$ and $\Delta_y$, shift the apex of the Dirac cones from the origin to $(-\Delta_y/\hbar v_\mathsmaller{\mathrm{F}},\Delta_x/\hbar v_\mathsmaller{\mathrm{F}})^T$ in the $k_xk_y$-plane. (ii) However, if $\bm{n}$ contains a component perpendicular to the TI surface, that is, $\Delta_z\neq0$, a gap in the spectrum of the TI surface states is opened.

A peculiar feature of TI surface states is the correlation between their spin and momentum, or more generally between their spin and velocity expectation values. Without an induced ferromagnetic exchange splitting $\bm{\Delta}$, the spin (expectation value) $\braket{\bm{s}}=(\hbar/2)\braket{\bm{\sigma}}$ and momentum $\hbar\bm{k}=(\hbar k_x,\hbar k_y)^T$ of TI surface states are locked, that is, they both have only in-plane components parallel to the surface and are orthogonal to each other: In the upper Dirac cone, $E>V_0$ (red in Fig.~\ref{fig:SuppDiracCones}), the angle between $\bm{k}$ and $\braket{\bm{s}}$ is always $\pi/2$, while in the lower Dirac cone, $E<V_0$ (blue in Fig.~\ref{fig:SuppDiracCones}), this angle is always $-\pi/2$ [see Figs.~\ref{fig:SuppDiracCones}~(a) and~(b)]. If $\bm{\Delta}\neq\bm{0}$, however, this correlation is no longer satisfied and the angle between $\bm{k}$ and $\braket{\bm{s}}$ is no longer fixed, but varies with $\bm{k}$.

Nevertheless, because the velocity operator for 3D TI surface states obtained from Eq.~(1) in the main text is
\begin{equation}\label{VelocityOperator}
\hat{\bm{v}}=v_\mathsmaller{\mathrm{F}}(\bm{e}_z\times\bm{\sigma}),
\end{equation}
the velocity expectation value $\braket{\hat{\bm{v}}}$ of---or equivalently the current density carried by---an eigenstate with momentum $\hbar\bm{k}$ is always in-plane and perpendicular to its spin expectation value $\braket{\bm{s}}$ even for finite $\bm{\Delta}$. We find the spin and velocity expectation values to be
\begin{equation}\label{SpinExpectation}
\braket{\bm{s}}=\frac{\hbar/2}{E-V_0-\Delta_z}\left(\begin{array}{c}
   \hbar v_\mathsmaller{\mathrm{F}}k_y-\Delta_x\\
   -(\hbar v_\mathsmaller{\mathrm{F}}k_x+\Delta_y)\\
   -\Delta_z\\
  \end{array}\right)
\end{equation}
and
\begin{equation}\label{VelocityExpectation}
\braket{\hat{\bm{v}}}=\frac{v_\mathsmaller{\mathrm{F}}}{E-V_0-\Delta_z}\left(\begin{array}{c}
   \hbar v_\mathsmaller{\mathrm{F}}k_x+\Delta_y\\
   \hbar v_\mathsmaller{\mathrm{F}}k_y-\Delta_x\\
   0\\
  \end{array}\right),
\end{equation}
where $E$ is given by Eq.~(\ref{Dispersion}). Hence, for eigenstates with $\Delta_z=0$, $\braket{\bm{s}}$ is also in-plane and, moreover, running clockwise (counterclockwise) and tangentially around the Fermi circle in the upper (lower) Dirac cone as depicted in Figs.~\ref{fig:SuppDiracCones}~(a) and~(b) here and Fig.~1~(c) of the main text. For eigenstates with $\Delta_z\neq0$, $\braket{\bm{s}}$ acquires also an out-of-plane component as displayed in Fig.~\ref{fig:SuppDiracCones}~(c). In both cases, $\braket{\hat{\bm{v}}}$ is aligned either parallel or anti-parallel to the radial direction of the Fermi circle (see Fig.~\ref{fig:SuppDiracCones}). Thus, whereas there is no longer a spin-momentum locking for $\bm{\Delta}\neq\bm{0}$, the more general spin-velocity locking remains also in this case.

\begin{figure}[t]
\centering
\includegraphics*[trim=0cm 8.25cm 4.5cm 9.5cm,clip,width=8.6cm]{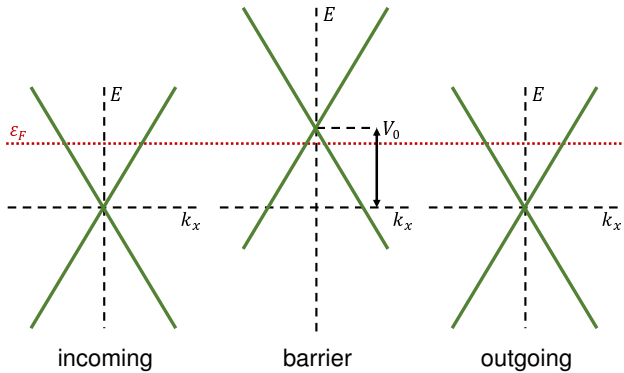}
\caption{(Color online) Scheme of the low-energy excitation spectra of a 3D-TI-based tunnel junction at $k_y=0$ and tunneling through a potential barrier of height $V_0$. The dotted line denotes the Fermi energy $\varepsilon_F$.}\label{fig:SuppKlein}
\end{figure}

Finally, we address another crucial aspect of 3D TI surface states. Since Eq.~(\ref{Dispersion}) describes a dispersion, which yields positive as well as negative energies with respect to the neutrality point as shown in Fig.~\ref{fig:SuppDiracCones}, we find that relativistic Klein tunneling, also observed in other Dirac-like systems, such as graphene \cite{Katsnelson2006:NP}, plays an important role that is affected by the ferromagnetic exchange. Originally, the Klein paradox refers to the counterintuitive phenomenon that, unlike in non-relativistic quantum mechanics, the transmission of a relativistic electron through a potential barrier does not decay exponentially with increasing barrier height $V_0$ (as long as $V_0$ exceeds the electronic rest energy which vanishes in our case).

\begin{figure}[t]
\centering
\includegraphics*[trim=0cm 8.25cm 4.5cm 9.5cm,clip,width=8.6cm]{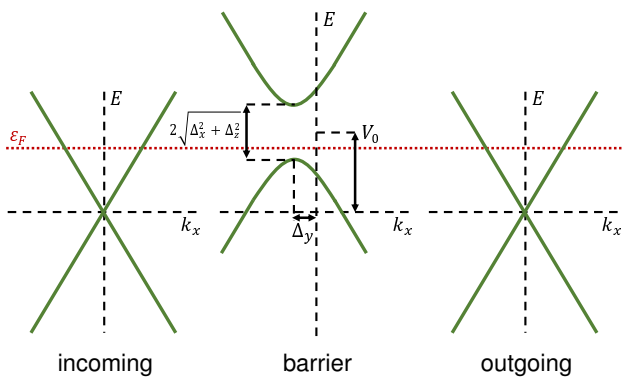}
\caption{(Color online) Scheme of the low-energy excitation spectra of a 3D-TI-based tunnel junction at $k_y=0$ and tunneling through a potential barrier of height $V_0$ with a ferromagnetic exchange splitting $2\sqrt{\Delta^2_x+\Delta^2_z}$ at $k_y=0$. The dotted line denotes the Fermi energy $\varepsilon_F$.}\label{fig:SuppKleinExchange}
\end{figure}

For illustration, Figs.~\ref{fig:SuppKlein} and~\ref{fig:SuppKleinExchange} show the schematic setups of the tunneling geometry with the dispersion of the isolated leads and barrier, each given by Eq.~(\ref{Dispersion}). In the absence of any ferromagnetic exchange field, Fig.~\ref{fig:SuppKlein} illustrates that at any (Fermi) energy---even for high barriers---there are always states available in the barrier through which an electron can tunnel, hence the absence of an exponential decay of the wave function in the barrier.

A finite $\bm{\Delta}$ in the barrier region can have profound consequences for tunneling through that barrier. While a finite $\Delta_z$ opens up of a gap between the conduction and valence states, the shift of the dispersion due to $\Delta_x$ restricts the phase space for Klein tunneling. The energy spectra~(\ref{Dispersion}) at $k_y=0$ are shown in Fig.~\ref{fig:SuppKleinExchange} and reflect that, as a consequence of both mechanisms, in the effective gap of $|E-V_0|<\sqrt{\Delta_x^2+\Delta_z^2}$ the transmission is strongly suppressed, while $\Delta_y$ shifts the momentum in the $x$-direction, but does not affect transport along the $x$-direction. In that sense, an induced exchange splitting by a ferromagnet with magnetization in the $x$- and $z$-directions has the same effect as the mass term in the relativistic Dirac equation: The wave functions of states with energies inside this effective gap decay exponentially inside the barrier like in non-relativistic quantum mechanics.

\section{Boundary conditions}\label{Sec:AppendixDelta}

The state $\Phi(x)$ given by Eq.~(2) in the main text is determined by invoking the boundary conditions at the interfaces between the leads and the barrier. These require the continuity of the wave function, $\Phi(0^+)=\Phi(0^-)$ and $\Phi(-d^+)=\Phi(-d^-)$, in the case of a finite barrier.

To obtain the boundary conditions for a $\delta$-barrier, we have to consider the Hamiltonian given by Eq.~(1) in the main text. With the ansatz $\Psi_{k_y}(x,y)=\exp(\i k_yy)\Phi(x)/\sqrt{2S}$ for the eigenstates, $\Phi(x)$ is determined by the differential equation
\begin{equation}\label{AppDelta_DiffEq}
\frac{\partial\Phi(x)}{\partial x}=\hat{\mathcal{D}}(x)\Phi(x)
\end{equation}
with
\begin{equation}\label{AppDelta_DiffOp}
\hat{\mathcal{D}}(x)=\frac{\i\sigma_y}{\hbar v_\mathsmaller{\mathrm{F}}}\left[\hbar v_\mathsmaller{\mathrm{F}}k_y\sigma_x-E+\left(V_0-\bm{\Delta}\cdot\bm{\sigma}\right)d\delta(x)\right].
\end{equation}

If the wave function at $x=x_0$ is given by $\Phi(x_0)$, Eq.~(\ref{AppDelta_DiffEq}) is solved by
\begin{equation}\label{AppDelta_DiffEqSolGen}
\Phi(x)=\exp\left[\int\limits_{x_0}^{x}\d x'\hat{\mathcal{D}}(x')\right]\Phi(x_0)
\end{equation}
for arbitrary $x$. If $x$ is chosen as an infinitesimally small number approaching zero from above and $x_0$ as an infinitesimally small number approaching zero from below, we obtain after integration
\begin{equation}\label{AppDelta_Propagator}
\begin{aligned}
\hat{U}&=\lim\limits_{\varepsilon\to0}\left\{\exp\left[\int\limits_{-\varepsilon}^{\varepsilon}\d x'\hat{\mathcal{D}}(x')\right]\right\}\\
&=\exp\left\{\frac{\i d}{\hbar v_\mathsmaller{\mathrm{F}}}\left[-\Delta_y+V_0 \sigma_y -\i \left(\bm{\Delta}\times\bm{\sigma}\right)\cdot\mathbf{e}_y\right]\right\},
\end{aligned}
\end{equation}
connecting the parts of the wave function for $x<0$ and for $x>0$. This is consistent with Ref.~\onlinecite{MatosAbiague2003:PRA}, where the temporal propagation of a hydrogen atom driven by electric-field pulses has been clarified.

Using the identity
\begin{equation}\label{AppDelta_Exp}
\exp\left[\i\alpha\bm{v}\cdot\bm{\sigma}\right]=\cos\left(\alpha\sqrt{\bm{v}^2}\right)+\i\frac{\bm{v}\cdot\bm{\sigma}}{\sqrt{\bm{v}^2}}\sin\left(\alpha\sqrt{\bm{v}^2}\right),
\end{equation}
valid for a complex number $\alpha$ and a complex vector $\bm{v}$, $\hat{U}$ can be rewritten as
\begin{equation}\label{AppDelta_Matrix}
\begin{aligned}
\hat{U}=\e^{-\i Z_y}\left(\begin{array}{ll} \cos Z-\frac{\Delta_x\sin Z}{\sqrt{V_0^2-\Delta^2}} & \frac{\left(\Delta_z+V_0\right)\sin Z}{\sqrt{V_0^2-\Delta^2}}\\ \frac{\left(\Delta_z-V_0\right)\sin Z}{\sqrt{V_0^2-\Delta^2}} & \cos Z+\frac{\Delta_x\sin Z}{\sqrt{V_0^2-\Delta^2}}\end{array}\right),
\end{aligned}
\end{equation}
where $\Delta=\sqrt{\Delta^2_x+\Delta^2_z}$. We emphasize that Eq.~(\ref{AppDelta_Exp}) is valid for any complex vector $\bm{v}$, even for $\sqrt{\bm{v}^2}=0$, when Eq.~(\ref{AppDelta_Exp}) reduces to
\begin{equation}\label{AppDelta_Exp0}
\exp\left(\i\alpha\bm{v}\cdot\bm{\sigma}\right)=1+\i\alpha\bm{v}\cdot\bm{\sigma}.
\end{equation}

\section{Tunneling current and conductance}\label{Sec:AppendixConductance}
\subsection{Low-bias current and linear conductance}
In order to calculate the tunneling conductance, we first compute the current due to an applied bias voltage along the $x$-direction. The charge current in the $i$-direction is given by
\begin{equation}\label{Current_General}
\begin{aligned}
I_i=eD_i\sum\limits_{\pm,k_x,k_y}\left\{\right.&j^{\mathsmaller{\mathrm{l}\to\mathrm{r}}}_i f_\mathrm{l}(E)\left[1-f_\mathrm{r}(E)\right]\\
&\left.+j^{\mathsmaller{\mathrm{r}\to\mathrm{l}}}_i f_\mathrm{r}(E)\left[1-f_\mathrm{l}(E)\right]\right\},
\end{aligned}
\end{equation}
where $j^{\mathsmaller{\mathrm{l}\to\mathrm{r}}}_i$ and $j^{\mathsmaller{\mathrm{r}\to\mathrm{l}}}_i$ denote the average particle current densities of particles moving to the right and left, respectively, while $f_\mathrm{l}(E)=f_\mathsmaller{\mathrm{FD}}(E-eV)$ and $f_\mathrm{r}(E)=f_\mathsmaller{\mathrm{FD}}(E)$ denote the distribution functions of the left and right reservoirs at energy $E=\pm E(k_x,k_y)$ with the Fermi-Dirac distribution function $f_\mathsmaller{\mathrm{FD}}(E)=1/\left\{\exp\left[\left(E-\mu\right)/\left(k_\mathrm{B}T\right)\right]+1\right\}$, the temperature $T$, the Boltzmann constant $k_\mathrm{B}$, and the chemical potential $\mu$. Here, the summation runs over the momenta in the $x$- and $y$-directions, $k_x$ and $k_y$, as well as over the conduction ($+$) and valence bands ($-$).

Since $j^{\mathsmaller{\mathrm{l}\to\mathrm{r}}}_i=-j^{\mathsmaller{\mathrm{r}\to\mathrm{l}}}_i\equiv j_i$, Eq.~(\ref{Current_General}) reduces to
\begin{equation}\label{Current_General_Simplified}
I_i=eD_i\sum\limits_{\pm,k_x,k_y}j_i\left[f_\mathrm{l}(E)-f_\mathrm{r}(E)\right].
\end{equation}
Here, the particle current density as derived from the Hamiltonian~(1) reads as
\begin{equation}\label{CurrentDensity_General}
\bm{j}=v_\mathsmaller{\mathrm{F}}\Psi^\dagger(x,y)\left(\bm{e}_z\times\bm{\sigma}\right)\Psi(x,y)
\end{equation}
for any given two-component spinor wave function $\Psi(x,y)$ [compare to the velocity operator given by Eq.~(\ref{VelocityOperator})].

Inserting the scattering states~(2) into Eq.~(\ref{CurrentDensity_General}) and evaluating $j_i$ at $x>0$ yields
\begin{equation}\label{Current_xx_Final}
\begin{aligned}
I_x=\frac{e}{h}\frac{D_x}{2\pi\hbar v_\mathsmaller{\mathrm{F}}}\int\limits_{-\infty}^{\infty}\d E\,|E|&\int\limits_{-\pi/2}^{\pi/2}\d\theta\cos\theta\;T(E,\theta)\\
&\times\left[f_\mathsmaller{\mathrm{FD}}(E-eV)-f_\mathsmaller{\mathrm{FD}}(E)\right],
\end{aligned}
\end{equation}
and
\begin{equation}\label{Current_yx_Final}
\begin{aligned}
I_y=\frac{e}{h}\frac{D_y}{2\pi\hbar v_\mathsmaller{\mathrm{F}}}\int\limits_{-\infty}^{\infty}\d E\,E&\int\limits_{-\pi/2}^{\pi/2}\d\theta\sin\theta\;T(E,\theta)\\
&\times\left[f_\mathsmaller{\mathrm{FD}}(E-eV)-f_\mathsmaller{\mathrm{FD}}(E)\right],
\end{aligned}
\end{equation}
where $h=2\pi\hbar$. In order to obtain Eqs.~(\ref{Current_xx_Final}) and~(\ref{Current_yx_Final}), which are valid for low bias $|eV|\ll|\varepsilon_\mathrm{F}|$, we have replaced the summations over $k_x$ and $k_y$ by
\begin{equation}\label{Summation}
\sum\limits_{\pm,k_x,k_y}=\frac{S}{\left(2\pi\hbar v_\mathsmaller{\mathrm{F}}\right)^2}\int\limits_{-\infty}^{\infty}\d E\,|E|\int\limits_{-\pi/2}^{\pi/2}\d\theta.
\end{equation}

By expanding Eqs.~(\ref{Current_xx_Final}) and~(\ref{Current_yx_Final}) up to the first order in the bias voltage $V$ and integrating the energy over the resulting $\delta$-function $T(E,\theta)\left[f_\mathsmaller{\mathrm{FD}}(E-eV)-f_\mathsmaller{\mathrm{FD}}(E)\right]\approx T(E,\theta)\delta(E-\varepsilon_\mathrm{F})eV$, one can calculate the conductances $G_{xx/yx}=I_{x/y}/V$ at zero temperature, given by Eq.~(6). Here, $T(E,\theta)$ is independent of the bias as in Eq.~(4).

Equation~(6) in the main text refers to conductances measured after the barrier, $x>0$. Similar expressions can be derived in front of the barrier, $x<-d$, where reflection also results in a transverse current. Using the expressions from Eq.~(6), the longitudinal conductance is given by $G_{xx}$, while the transverse conductance is given by $-G_{yx}$. Likewise, the longitudinal conductance measured inside the barrier is also given by $G_{xx}$ from Eq.~(6), whereas the net transverse conductance measured across the entire barrier region, $-d<x<0$, vanishes. For expressions of the Hall voltages and resistances, we refer to Sec.~\ref{Sec:HallR&V}.

\begin{figure}[t]
\centering
\includegraphics*[trim=0.5cm 1.35cm 0.5cm 14.0cm,clip,width=8.6cm]{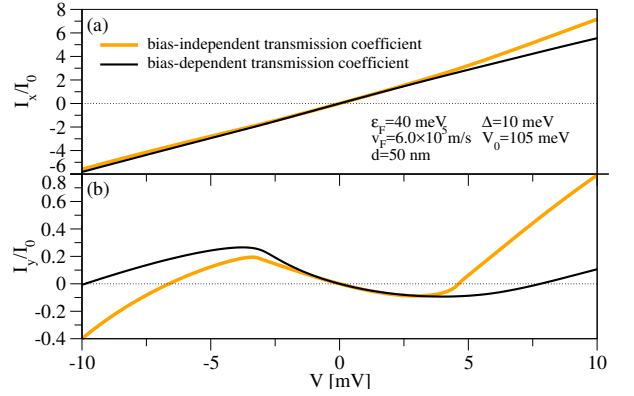}
\caption{(Color online) Comparison between the (a) longitudinal and (b) TPH currents of a finite barrier in the low-bias regime, $|eV|\ll\varepsilon_\mathsmaller{\mathrm{F}}$, if a bias-independent and a bias-dependent transmission is used. The magnetization points along the $x$-direction.}\label{fig:SuppBias}
\end{figure}

\begin{figure}[t]
\centering
\includegraphics*[trim=0.5cm 1.35cm 0.5cm 14.0cm,clip,width=8.6cm]{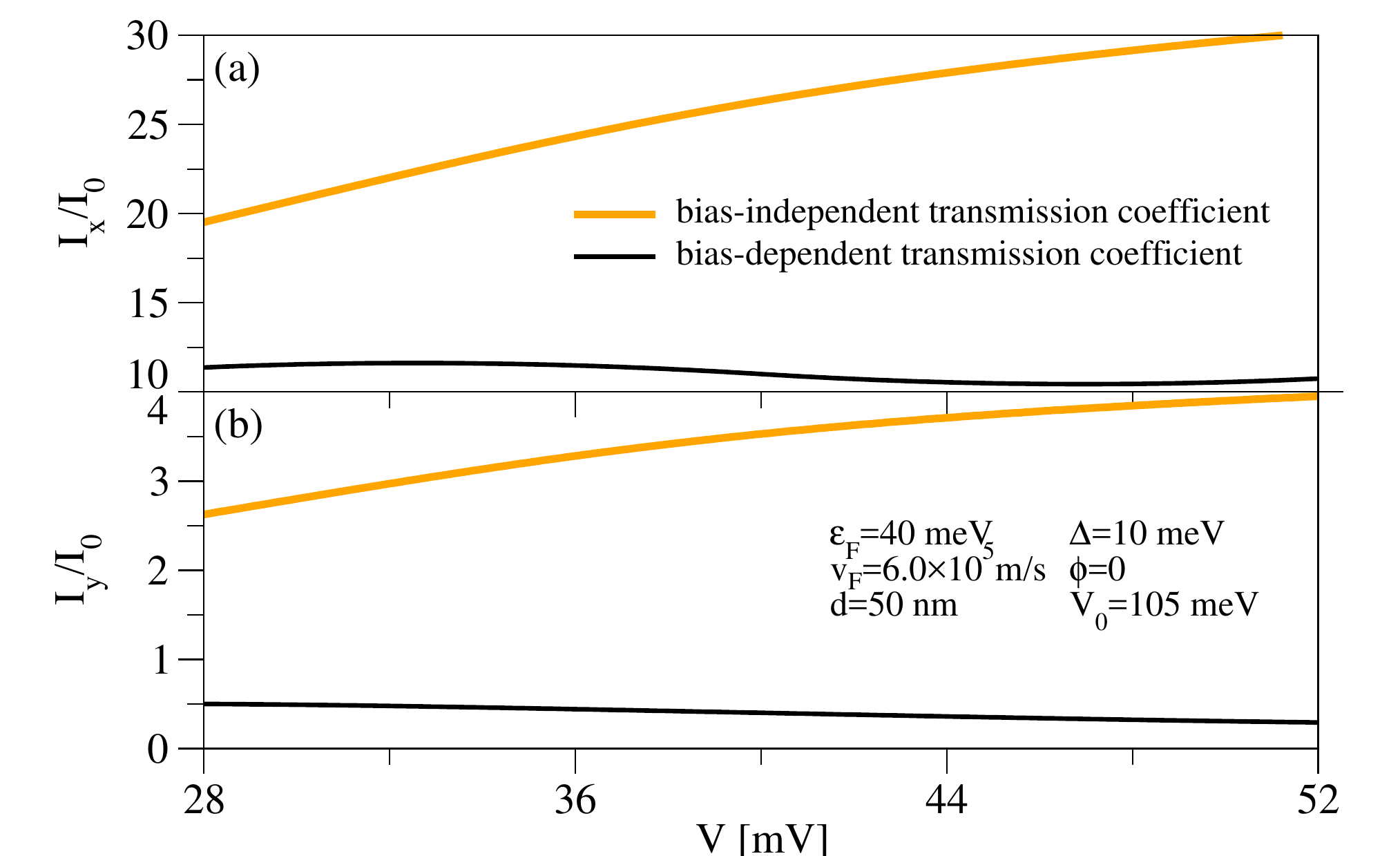}
\caption{(Color online) Comparison between the (a) longitudinal and (b) TPH currents of a finite barrier in the high-bias regime, $|eV|\gtrsim\varepsilon_\mathsmaller{\mathrm{F}}$, if a bias-independent and a bias-dependent transmission is used. The magnetization points along the $x$-direction.}\label{fig:SuppBias_High}
\end{figure}

\begin{figure}[t]
\centering
\includegraphics*[trim=0.5cm 1.35cm 0.5cm 14.0cm,clip,width=8.6cm]{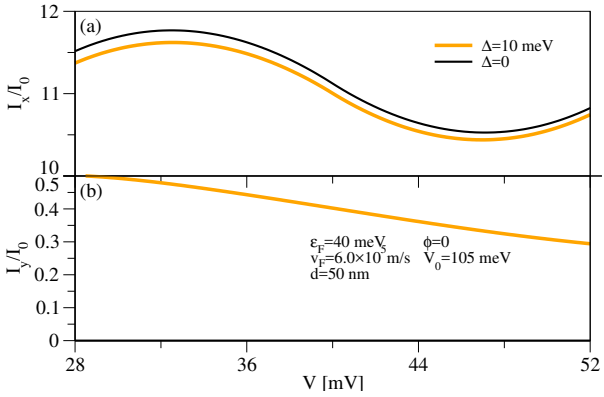}
\caption{(Color online) Comparison between the (a) longitudinal and (b) TPH currents of a finite non-magnetic barrier in the high-bias regime, $|eV|\gtrsim\varepsilon_\mathsmaller{\mathrm{F}}$, if a bias-independent and a bias-dependent transmission is used.}\label{fig:SuppBias_NoMagneticBarrier}
\end{figure}

\subsection{High-bias current and differential conductance}
Unlike those conductances and the low-bias currents~(\ref{Current_xx_Final}) and~(\ref{Current_yx_Final}), currents at high bias, $|eV|\gtrsim\varepsilon_\mathsmaller{\mathrm{F}}$, and their corresponding differential conductances $G_{xx/yx}(V)=\d I_{x/y}(V)/\d V$ need to be calculated by taking into account the bias dependence of the transmission $T(E,\theta)$ and the velocity mismatch between the left and right leads due to bias-induced changes of the potential profile. We model this effect by adding the step-like~\cite{Tsu1973:APL} potential profile $V\left[\Theta(-x-d)+\Theta(-x)\right]/2$ to the Hamiltonian~(1) in the main text, solving the scattering problem for this system and computing the transmitted currents numerically. However, for large Fermi energies, $|\varepsilon_\mathrm{F}|\gg|eV|$, the effect of the bias dependence on $T(E,\theta)$ and the Fermi velocity mismatch is weak and one can approximately treat $T(E,\theta)$ by evaluating it at $V=0$ as done in Eq.~(4). Then, the differential conductances $G_{xx/yx}(V)$ can be computed as given by Eq.~(6) if $\varepsilon_\mathrm{F}$ is replaced by $\varepsilon_\mathrm{F}+eV$. Figure~\ref{fig:SuppBias} compares the calculated currents for the setup of Fig.~4 in the main text if the bias dependence of $T(E,\theta)$ is not taken into account and if an additional step-like potential due to the bias voltage $V$ is included in the calculation of $T(E,\theta)$. For low bias, $|eV|\ll\varepsilon_\mathsmaller{\mathrm{F}}$, the qualitative behavior of the currents is the same in both cases and, in contrast to a ND longitudinal conductance, the NDHC appears even without considering electrostatic effects in $T(E,\theta)$.

\begin{figure}[t]
\centering
\includegraphics*[trim=0.5cm 1.35cm 0.5cm 14.0cm,clip,width=8.6cm]{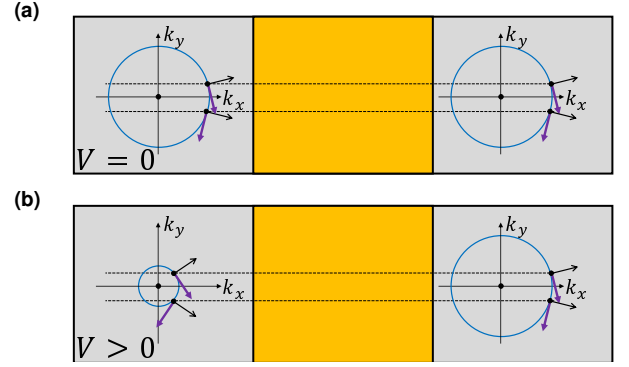}
\caption{(Color online) Origin of the ND longitudinal conductance at high bias, $|eV|\gtrsim\varepsilon_\mathsmaller{\mathrm{F}}$, due to the spin/velocity mismatch between the leads: Fermi circles in the TI for $\varepsilon_\mathsmaller{\mathrm{F}}>0$ and (a) $V=0$ and (b) $V>0$. Here, violet (black) arrows denote the electron spin orientation (direction of motion).}\label{fig:SuppBias_Mismatch}
\end{figure}

On the other hand, the ND longitudinal conductance in the high-bias regime shown in Fig.~\ref{fig:SuppBias_High} is due to the bias-induced potential profile, but also appears in the absence of a magnetic barrier (Fig.~\ref{fig:SuppBias_NoMagneticBarrier}). It is thus a very different effect than the NDHC we predict at low bias that crucially depends on the presence of a magnetic barrier. The ND longitudinal conductance is caused by the mismatch between the Fermi circles in the two leads: As can be seen in Fig.~\ref{fig:SuppBias_Mismatch}, a finite bias voltage between the two leads results in Fermi circles of different radii, $|\varepsilon_\mathsmaller{\mathrm{F}}-eV|$ (left lead in our setup) and $|\varepsilon_\mathsmaller{\mathrm{F}}|$ (right lead in our setup), and hence a spin/velocity mismatch between the leads. Including the step-like potential profile, one can write the current as
\begin{equation}\label{SuppV_Current}
\begin{aligned}
I_{x/y}=\frac{e}{h}\frac{D_{x/y}}{2\pi\hbar v_\mathsmaller{\mathrm{F}}}&\int\limits_{-\infty}^{\infty}\d E\int\limits_{-\pi/2}^{\pi/2}\d\theta\,T_{x/y}(E,\theta,V)\\
&\times\left[f_\mathsmaller{\mathrm{FD}}(E-eV)-f_\mathsmaller{\mathrm{FD}}(E)\right],
\end{aligned}
\end{equation}
where
\begin{equation}\label{SuppV_Tx}
\begin{aligned}
T_x(E,\theta,V)=&\frac{|E-eV|}{|E|}\mathrm{Re}\left[\sqrt{E^2-(E-eV)^2\sin^2\theta}\right]\\
&\times|t_\mathrm{e}(E,\theta,V)|^2
\end{aligned}
\end{equation}
and
\begin{equation}\label{SuppV_Ty}
\begin{aligned}
T_y(E,\theta,V)=&\frac{(E-eV)^2\sin\theta}{E}\Theta\left[E^2-(E-eV)^2\sin^2\theta\right]\\
&\times|t_\mathrm{e}(E,\theta,V)|^2
\end{aligned}
\end{equation}
contain the transmission $T(E,\theta,V)=|t_\mathrm{e}(E,\theta,V)|^2$ and the velocity mismatch between the left and right leads. By taking the limit $V=0$ in Eqs.~(\ref{SuppV_Tx}) and~(\ref{SuppV_Ty}) and inserting them into Eq.~(\ref{SuppV_Current}), one can recover the low-bias expressions~(\ref{Current_xx_Final}) and~(\ref{Current_yx_Final}) above. The differential conductance $G_{xx}(V)$ then consists of two terms:
\begin{equation}\label{SuppV_Conductance}
\begin{array}{l}
G_{xx}(V)=\frac{e}{h}\frac{D_x}{2\pi\hbar v_\mathsmaller{\mathrm{F}}}\times\\
\left\{\int\limits_{-\infty}^{\infty}\d E\int\limits_{-\pi/2}^{\pi/2}\d\theta\;\frac{\d T_x(E,\theta,V)}{\d V}\left[f_\mathsmaller{\mathrm{FD}}(E-eV)-f_\mathsmaller{\mathrm{FD}}(E)\right]\right.\\
\left.+e\int\limits_{-\infty}^{\infty}\d E\int\limits_{-\pi/2}^{\pi/2}\d\theta\;T_x(E,\theta,V)\left[-\frac{\d f_\mathrm{FD}(E-eV)}{\d E}\right]\right\}.
\end{array}
\end{equation}
Omitting the bias dependence of $T_x(E,\theta,V)$, the first term in Eq.~(\ref{SuppV_Conductance}) vanishes and one obtains a differential conductance $G_{xx}(V)$ that is always positive. This is because $T(E,\theta,V)$ as well as the velocity in the $x$-direction and hence $T_x(E,\theta,V)$ in the second term in Eq.~(\ref{SuppV_Conductance}) are always positive. If the bias dependence of $T_x(E,\theta,V)$ is taken into account, $\d T_x(E,\theta,V)/\d V<0$ due to the spin/velocity mismatch between the leads illustrated in Fig.~\ref{fig:SuppBias_Mismatch} and there is a competition between the two contributions in Eq.~(\ref{SuppV_Conductance}). Typically the second (that is, positive) contribution to $G_{xx}(V)$ dominates. Close to $eV\approx\varepsilon_\mathsmaller{\mathrm{F}}$, however, this second contribution is very small and actually vanishes for $eV=\varepsilon_\mathsmaller{\mathrm{F}}$ at zero temperature. Then Eq.~(\ref{SuppV_Conductance}) above yields negative values and a ND longitudinal conductance. In Fig.~4 in the main text, we have included all those effects due to $V$.

\section{Dependence on the direction of the magnetization}\label{Sec:AppendixDirection}

\begin{figure}[t]
\centering
\includegraphics*[trim=0cm 9.5cm 4.25cm 6.25cm,clip,width=8.6cm]{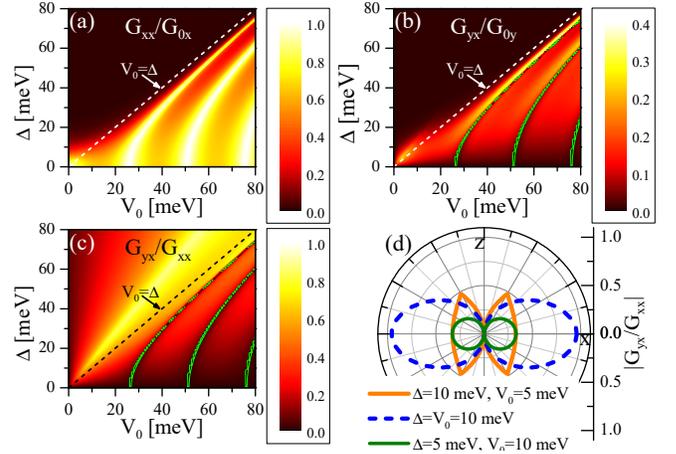}
\caption{(Color online) Dependence of the (a) longitudinal and (b) transverse conductances as well as (c) of their ratio on $V_0$ and $\Delta$ for a finite barrier with $d=50$ nm, $n_y=0$, $\phi=\pi/4$ between $n_x$ and $n_z$, $\varepsilon_\mathrm{F}=1$ meV, $v_\mathrm{F}=6.0\times10^5$ m/s. Panel~(d) displays a polar plot of the (absolute value of the) ratio shown in panel~(c) with respect to the angle $\phi$ between $n_x$ and $n_z$ for different $V_0$ and $\Delta$ (for $n_y=0$). Green lines denote the boundaries of regions with negative conductance.}\label{fig:SuppPolar}
\end{figure}

The induced ferromagnetic exchange field $\bm{\Delta}=|\bm{\Delta}|\bm{n}\equiv(\Delta_x,\Delta_y,\Delta_z)^T$ depends on the magnetization orientation of the ferromagnet on top of the TI given by the unit vector $\bm{n}$ (the energy gap between the two split bands is given by $2|\bm{\Delta}|$). Thus, the quantity $\Delta=\sqrt{\Delta_x^2+\Delta_z^2}=|\bm{\Delta}|\sqrt{n_x^2+n_z^2}$ is given by the projection of $\bm{n}$ into the $xz$-plane multiplied by the exchange splitting. Whilst they do not depend on $\Delta_y$, $G_{xx}$ and especially $G_{yx}$ exhibit a marked behavior with the direction of the magnetization in the $xz$-plane described by the angle $\phi$ between $n_x=\sqrt{n_x^2+n_z^2}\cos\phi$ and $n_z=\sqrt{n_x^2+n_z^2}\sin\phi$ as demonstrated in Fig.~\ref{fig:SuppPolar}. For convenience, we have set $n_y=0$ here, implying $\Delta=|\bm{\Delta}|$.

The behavior of $G_{yx}$ and $G_{yx}/G_{xx}$ differs between the regimes of Klein tunneling and tunneling with exponential decay, which can be observed in Fig.~\ref{fig:SuppPolar}~(d): For $|V_0|\geq\Delta$, $|G_{yx}/G_{xx}|$ is maximal if the magnetization is along the $x$-axis, but it is maximal for different magnetization orientations $\phi$ away from the $x$-axis if $|V_0|<\Delta$. This can be understood from $T(E,\theta)$ given by Eq.~(4) in the main text, where the factor
\begin{equation}\label{Prefactor}
\left[V_0\sgn(E)\sin\theta-\Delta_x\right]^2+\Delta_z^2\cos^2\theta
\end{equation}
introduces an asymmetry in the tunneling for opposite incident angles due to $\Delta_x$. If Eq.~(\ref{Prefactor}) vanishes for a given $\theta$, an electron with this incident angle is transmitted perfectly through the barrier, while electrons with opposite incident angles are much less likely to be transmitted, which in turn results in a relatively large value of $|G_{yx}/G_{xx}|$. This, however, is only possible if $\Delta_z=0$ and $|V_0|\geq|\Delta_x|$. Therefore, in the regime of Klein tunneling, $|G_{yx}/G_{xx}|$ is maximal for $\phi=0$, that is, $\Delta_z=0$, while finite values of $\phi$ result in a diminution of $|G_{yx}/G_{xx}|$. In the regime of decaying tunneling, Eq.~(\ref{Prefactor}) can never vanish completely, but can be minimized if $|V_0|\geq|\Delta_x|$ to yield a large transmission for certain incident angles compared to incident angles of opposite sign. Since $\Delta=\sqrt{\Delta_x^2+\Delta_z^2}\geq|V_0|$ in this regime, this means that $\Delta_z$ needs to be finite and that $|G_{yx}/G_{xx}|$ is maximal for magnetization orientations $\phi\neq0$. We remark that here we have ignored the resonance condition given by the second equality in Eq.~(5) in the main text, which can lead to additional modifications, such as the regions of negative Hall conductance shown in Figs.~\ref{fig:SuppPolar}~(b) and~(c). The origin of these regions is the same as in the case of $\Delta=|\Delta_x|$ discussed in the main text.

Figure~\ref{fig:SuppPolar} also illustrates the interplay between $V_0$ and $\Delta$ for both (a) $G_{xx}$ and (b) $G_{yx}$ as well as (c) their ratio for the same parameters as in Fig. 3 in the main text, but with an angle $\phi=\pi/4$ between $n_x$ and $n_z$. As before, Figs.~\ref{fig:SuppPolar}~(a) and~(b) clearly show the two different regions of Klein and increasingly suppressed tunneling, $|V_0|>\Delta$ and $|V_0|<\Delta$, respectively. However, the maximal ratios between the transverse and longitudinal tunneling conductances (assuming $D_x=D_y$) are shifted from $|V_0|\approx\Delta$ to the region of exponentially suppressed tunneling [see Fig.~\ref{fig:SuppPolar}~(c)].

At last, Fig.~\ref{fig:SuppPolar}~(d) illustrates again that in both regimes, the oscillatory regime and the regime of exponential decay, the transverse Hall signal vanishes for $\phi=\pm\pi/2$, that is, for a purely out-of-plane magnetization. This clearly distinguishes our effect from the AHE and the TAHE, where an out-of-plane magnetization is crucial for the emergence of a Hall signal.

\section{Tunneling planar Hall voltage and resistance}\label{Sec:HallR&V}
To give a transparent description of our calculations, we have presented expressions and plots for the tunneling conductances or currents in the main text. In experiments, however, one typically measures the Hall voltage, expressions for which are provided in the following. The currents and voltages are related by
\begin{equation}\label{i-g-v}
\left(\begin{array}{c} 
I_x \\
I_y \\
\end{array}\right)
= \left(\begin{array}{cc}
G_{xx} & G_{xy}\\
G_{yx} & G_{yy}
\end{array}\right)\left(\begin{array}{c}
V_x \\
V_y \\
\end{array}\right).
\end{equation}
Inverting the relation above, we obtain
\begin{equation}\label{v-r-i}
\left(\begin{array}{c} 
V_x \\
V_y \\
\end{array}\right)
= \left(\begin{array}{cc}
R_{xx} & R_{xy}\\
R_{yx} & R_{yy}
\end{array}\right)\left(\begin{array}{c}
I_x \\
I_y \\
\end{array}\right),
\end{equation}
where
\begin{equation}\label{r-long}
R_{xx}=\frac{G_{yy}}{\mathcal{D}},\;R_{yy}=\frac{G_{xx}}{\mathcal{D}},
\end{equation}
\begin{equation}\label{r-trans}
R_{xy}=-\frac{G_{xy}}{\mathcal{D}},\;R_{yx}=-\frac{G_{yx}}{\mathcal{D}},
\end{equation}
and
\begin{equation}
\mathcal{D}=G_{xx}G_{yy}-G_{xy}G_{yx}.
\end{equation}
The quantities characterizing the Hall response depend on the operation mode \cite{Popovic2004}.

\subsection{Hall voltage operation mode}
This is the more common mode, in which the output signal is the Hall voltage $V_\mathrm{H}=V_y$ under open circuit conditions in the $y$-direction ($I_y=0$) \cite{Popovic2004}. The Hall voltage in units of the longitudinal bias can be found by using Eqs.~(\ref{v-r-i})-(\ref{r-trans}),
\begin{equation}
\frac{V_\mathrm{H}}{V}=-\frac{G_{yx}}{G_{0y}}.
\end{equation}
Here, we have taken into account that, for the device under investigation, $V_x=V$ and $G_{yy}=G_{0y}$ is the Sharvin conductance given in the main text. Similarly, we have found the Hall resistance
\begin{equation}
R_\mathrm{H}=\frac{V_y}{I_x}=-\frac{G_{yx}}{\mathcal{D}}.
\end{equation}
In the limit $G_{xy}G_{yx}\ll G_{xx}G_{0y}$, that is, if the longitudinal conductance dominates, the Hall resistance reduces to
\begin{equation}
\frac{R_\mathrm{H}}{R_{0y}}\approx-\frac{G_{yx}}{G_{xx}}, 
\end{equation}
where $R_{0y}=1/G_{0y}$. On the other hand, in the giant TPHE regime,
\begin{equation}
R_{\rm H}\approx \frac{1}{G_{yx}}.
\end{equation}

\subsection{Hall current operation mode}
In this mode, the output signal is the Hall current $I_\mathrm{H}=I_{y}$, which produces a disturbance in the terminal currents \cite{Popovic2004}. For a closed circuit in the $y$-direction ($V_y=0$), the Hall current found from Eq.~(\ref{i-g-v}) is
\begin{equation}
I_\mathrm{H}= G_{yx}V_{x}
\end{equation}
with an associated Hall resistance~\cite{Popovic2004,Nachtwei1990:SST}
\begin{equation}
R_\mathrm{H}=\frac{V_x}{I_y}=\frac{1}{G_{yx}}.
\end{equation}
The corresponding Hall angle $\theta_{\rm H}$ is given by
\begin{equation}
\tan\theta_{\rm H}=\frac{I_{\rm H}}{I_x}=\frac{G_{yx}}{G_{xx}}.
\end{equation}

\begin{figure}[t]
\centering
\includegraphics*[trim=0.5cm 7.5cm 16.5cm 0.5cm,clip,width=8.0cm]{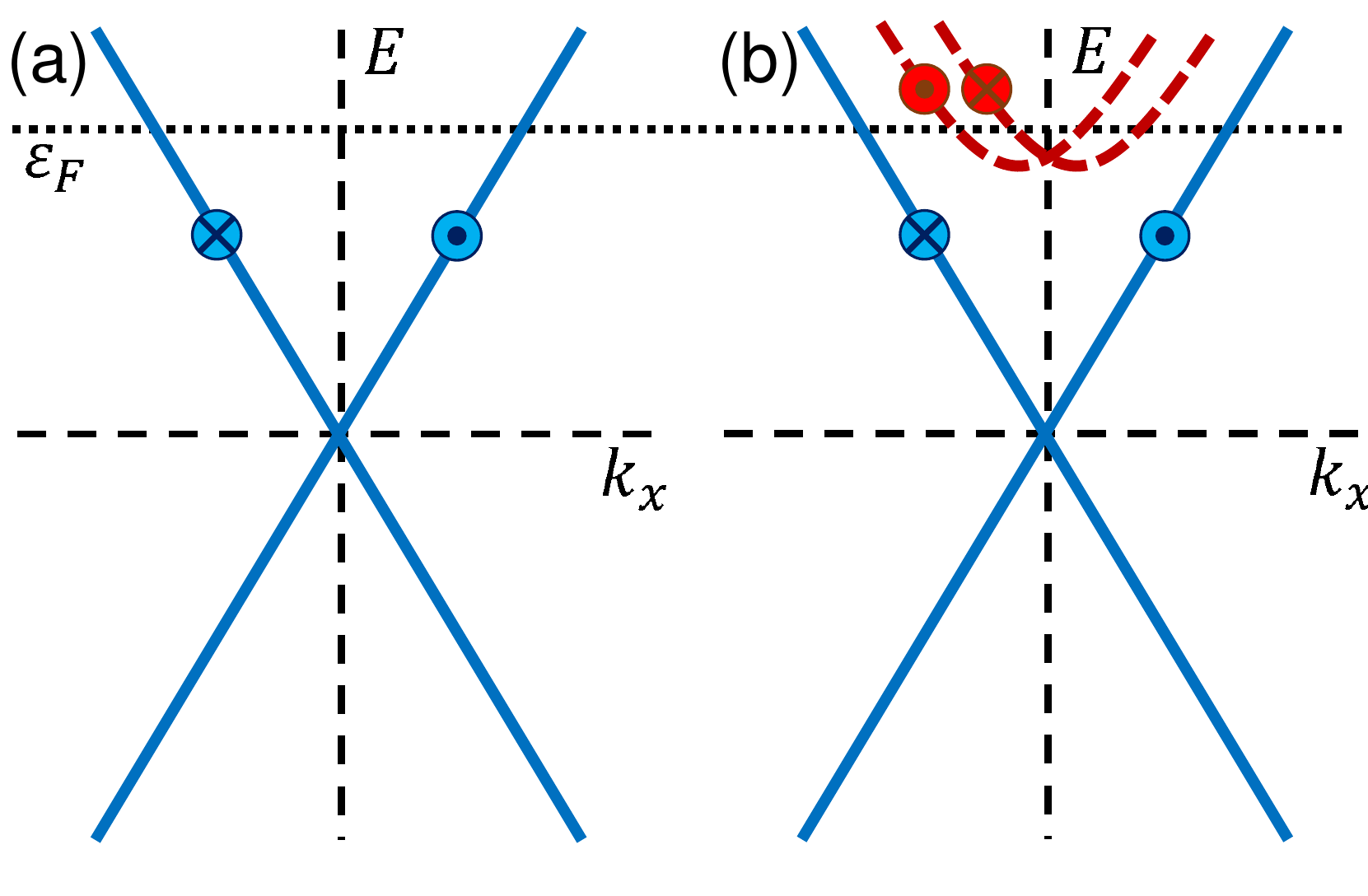}
\caption{(Color online) (a) TI surface states (TSSs) and (b) TSSs coexisting with Rashba 2D states (here without ferromagnetic exchange splitting). Here, the dots and crosses denote different spin helicities.}\label{fig:PureTIvsRashbaTI}
\end{figure}

\section{Topologically trivial states}\label{Sec:Materials}
The model used in this work considers only the Dirac-like topologically protected TI surface states (TSSs). However, it is known that even in the absence of magnetic proximity effects, these TSSs can coexist with other trivial states at the Fermi level due to band bending in certain TIs~\cite{Yang2016:PRB,Li2016:arxiv}. This situation is illustrated in Fig.~\ref{fig:PureTIvsRashbaTI}, which shows (a) only the TSSs, that is, the situation in our model, and (b) a situation where the TSSs coexist with Rashba 2D states at the Fermi level (both in the absence of magnetic proximity effects). For Bi$_2$Se$_3$ in proximity to magnetic materials, first-principles calculations suggest that there are no isolated TSSs at the Fermi level~\cite{Eremeev2013:PRB,*Lee2014:PRB2}.

Recent experiments point to (Bi$_x$Sb$_{1-x}$)$_2$Te$_3$/YIG heterostructures avoiding this problem and exhibiting isolated TSSs~\cite{Jiang2015:NL,*Jiang2016:NC}, that is, being a system described by our model. Nevertheless, we also briefly discuss the qualitative effect of additional Rashba 2D states on the TPHE in the following and argue that a transverse signal is also to be expected in this case.

\begin{figure}[t]
\centering
\includegraphics*[trim=0.5cm 7.5cm 17.0cm 0.5cm,clip,width=8.0cm]{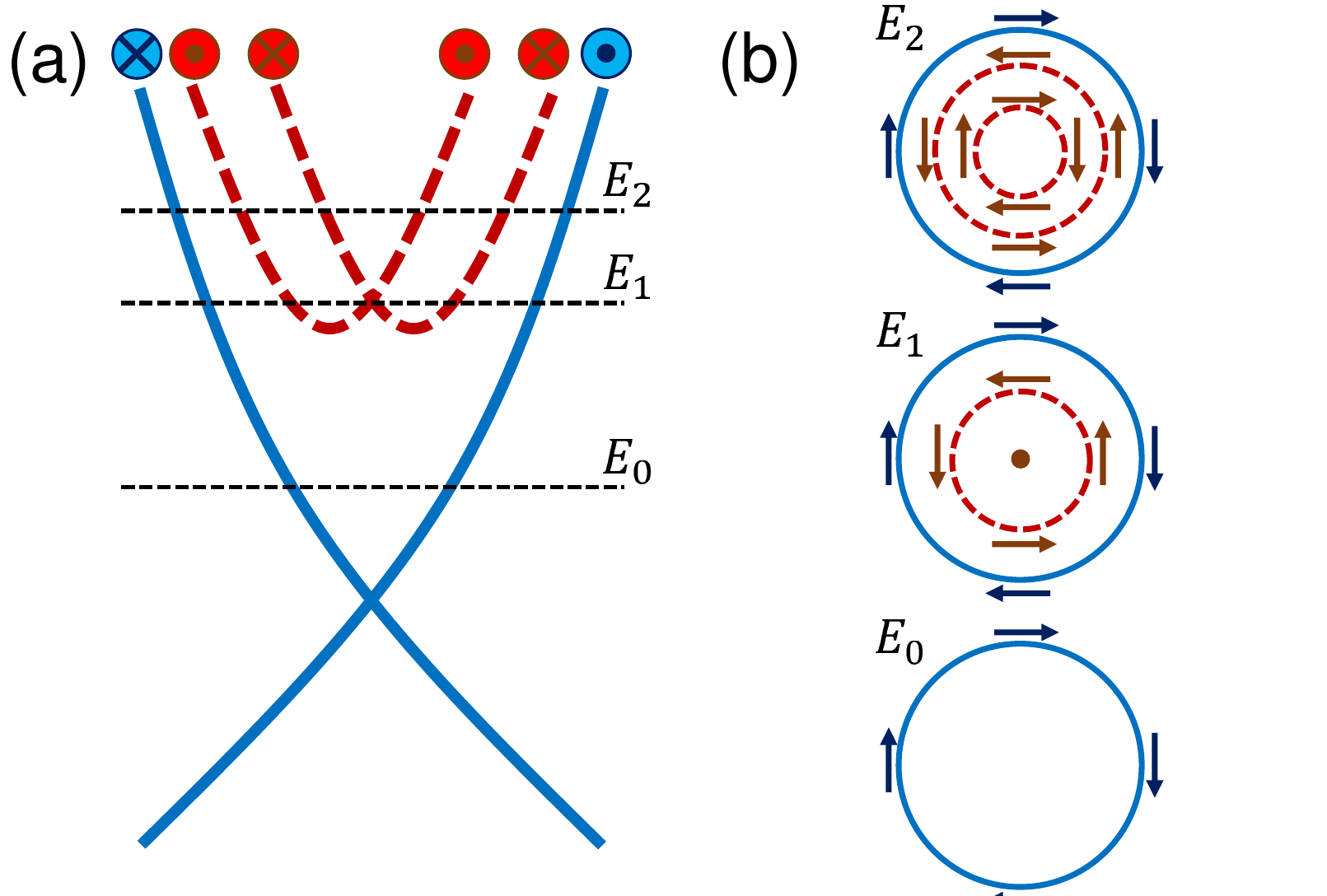}
\caption{(Color online) (a) Band structure and (b) Fermi contours at selected Fermi energies for TSSs and Rashba 2D states (here without ferromagnetic exchange splitting). Here, the dots, crosses, and arrows denote different spin helicities.}\label{fig:RashbaTI}
\end{figure}

If such spin-split Rashba 2D states are coexisting with the TSSs at the Fermi level, those states will in general also exhibit a spin mismatch and thus contribute to the transverse Hall voltage/conductance, potentially competing with the transverse Hall voltage due to the TSSs, as illustrated by Fig.~\ref{fig:PureTIvsRashbaTI}, where the band structure in the absence of magnetic proximity effects is shown (An in-plane magnetization would not only shift the Fermi circles, but also slightly deform the Fermi contours of the 2D Rashba states.). The TPHE originates from the presence of a finite net helicity in the system. If the Fermi energy lies inside the bulk gap [$E_0$ in Fig.~\ref{fig:PureTIvsRashbaTI}], as considered in this work, there is only one helicity and the TPHE, dominated by the TSSs, is large. As the Fermi energy increases, TSSs and Rashba 2D states coexist and start to compete. Depending on the respective Fermi velocities of the TSSs and Rashba 2D states, the TPHE can eventually be suppressed at $\varepsilon_\mathsmaller{\mathrm{F}}=E_1$ [$E_1$ in Fig.~\ref{fig:PureTIvsRashbaTI}], where the helicity of the Rashba states nearly cancels that of the TSSs. A finite TPHE is expected to reappear as the Fermi energy is further increased [$E_2$ in Fig.~\ref{fig:PureTIvsRashbaTI}]. In such a situation, the helicity of the outer Rashba states nearly cancels that of the TSSs and the TPHE signal is dominated by the inner Rashba states. Therefore, we expect the TPHE to be finite as long as the Fermi energy lies away from the Dirac point of the Rashba 2D states. For energies below, the TPHE is dominated by TSSs, while it can be dominated by Rashba states for energies above. Similar arguments have been used to interpret recent measurements of current-generated spin polarization due to spin-momentum locking either in TI Dirac surface states or trivial Rashba 2D states~\cite{Li2016:arxiv}, where the contribution to the total spin polarization can be tuned to be dominated by the TSSs~\cite{Li2016:arxiv}.

\end{document}